# Probing hot electron transport across an epitaxial Schottky interface of SrRuO$_3$/Nb:SrTiO$_3$

S. Roy, A. M. Kamerbeek, K. G. Rana, S. Parui and T. Banerjee*

Physics of Nanodevices, Zernike Institute for Advanced Materials,
University of Groningen, The Netherlands, 9747, AG
*E-mail: t.banerjee@rug.nl

**Abstract:** SrRuO$_3$ (SRO), a conducting transition metal oxide, is commonly used for engineering domains in BiFeO$_3$. New oxide devices can be envisioned by integrating SRO with an oxide semiconductor as Nb doped SrTiO$_3$ (Nb:STO). Using a three-terminal device configuration, we study vertical transport in a SRO/Nb:STO device at the nanoscale and find local differences in transport, that originate due to the high selectivity of SRO growth on the underlying surface terminations in Nb:STO. This causes a change in the interface energy band characteristics and is explained by the differences in the spatial distribution of the interface-dipoles at the local Schottky interface.

Complex oxide heterointerfaces are of immense interest in oxide electronics and has led to the emergence of new physical phenomena that arises due to competing energy scales involved in the interplay between the structural, charge, spin and orbital degrees of freedom.[1-4] Moreover, the possibility to control and manipulate carrier spin polarization at heterointerfaces between transition metal oxides [5-6] and the demonstration of novel functionalities at the domain walls in multiferroic materials [7-8] have opened new frontiers in oxide spintronics. A promising material among the complex oxides is SrRuO$_3$, which apart from being metallic is also ferromagnetic (below 160 K) [9] and thus can find use both in oxide electronics and spintronics. In spite of this, not much is known about electron transport across a functional interface of SRO particularly with an oxide semiconductor as Nb doped SrTiO$_3$, although the physical properties of SRO films have been extensively studied. It has been reported that the surface terminating plane of the underlying substrate can influence the initial growth of the deposited SRO films and has been proposed to be an interesting approach to fabricate ordered oxide nanostructures [10-11]. Trenches that are created in thin films of SRO are associated with a local surface termination of SrO at the STO substrate, where the initial growth of SRO is unfavorable. Regions in between the trenches are areas of

TiO$_2$ termination where SRO film grows unhindered. In this work, we use the technique of Ballistic Electron Emission Microscopy (BEEM) [12-14] and use its local probing capabilities and high spatial resolution, to investigate the differences in electron transport across SRO (8 unit cell) /Nb:STO, at regions where the local termination varies between SrO and TiO$_2$. Based on the transmission of hot electrons perpendicular to the epitaxial Schottky interface of SRO/Nb:STO, we observe variations in transmission using BEEM, between the two regions, the transmission being higher on the trenches (SrO termination) and corresponding to a larger Schottky barrier height (SBH). At the TiO$_2$ terminated regions, the transmission in SRO is lower with a corresponding low SBH. Electrical characterization of the same Schottky interface reveals an unique value of the SBH with good rectification at room temperature (RT). Using Scanning Tunnelling Spectroscopy (STS), we find local variations of the surface conductivity as manifested by the differences in the local density of surface states in SRO at the trenches (Region A) and outside them (Region B). To understand the differences in the local energy bands at the Schottky interface we invoke the concept of interface dipole strength and show that the nature and spatial distribution of the orbitals change at the differently terminated regions – a feature that can be resolved using the high spatial resolution and local probing capabilities of the BEEM. Thus using BEEM and combining it with STS studies, we show variations in the local conductance landscape both spatially as well at buried interfaces that arises due to the differences in the growth of SRO on Nb:STO and is a new route to study electron transport in complex oxide devices.

BEEM (Figure 1) employs the tip of a Scanning Tunnelling Microscope (STM) to inject hot electrons (energy a few electron volts above the Fermi level) across a vacuum-tunnelling barrier into a metallic over layer (base) which forms a Schottky interface with the semiconducting substrate (collector). The injected hot electron propagates perpendicular to the film plane, undergoes scattering in the base where the transmission exponentially decreases with the film thickness. [15] After transmission, a fraction of the hot electrons reach the Metal/Semiconductor (*M/S*) interface and are detected as BEEM current ($I_B$), provided the necessary energy and momentum criteria to overcome the Schottky Barrier Height (SBH) is satisfied at that interface. The collected BEEM current ($I_B$), gives us local information of hot electron transport at the Schottky interface, whereas the SBH at the nanoscale can be determined using the Bell-Kaiser (BK) model. [12]

Thin films of $SrRuO_3$ were deposited on (001) Nb doped $SrTiO_3$ substrates (0.01 wt% Nb doping) using a Pulsed Laser Deposition (PLD) system. It has been recognized that a clean and well defined substrate is imperative for the growth of uniform SRO, thus we treat the substrates using the standard chemical protocol followed by an annealing at 960°C in oxygen, for 1.5 hours. [16] This protocol is carried out to obtain a singly terminated $TiO_2$ surface. However, during the annealing process, Sr diffuses out to the steps resulting in occasional SrO terminations along with $TiO_2$ terminated regions on the substrates. The SRO films were grown using PLD with a KrF excimer laser at a repetition rate of 1 Hz at 600°C in an oxygen pressure of 0.13 mbar. The laser was focused at an incident angle of 45° to the target to obtain a fluence of around 2 J/cm$^2$. For this work, SRO films of 8 unit cells (u.c.; 1 u.c.= 0.39 nm) were deposited and the films were cooled down to RT in presence of oxygen at a pressure of 100 mbar.

The morphology of the Nb:$SrTiO_3$ substrate, after chemical treatment and annealing, is shown in the atomic force micrograph (AFM) image in Figure 2 (a) along with the step height corresponding to one unit cell of STO (white line in the image). Figure 2 (b) shows the AFM image of the 8 u.c. SRO film on the Nb:STO substrate. The AFM topography shows distinct terraces and from the height profile, a step height of 1 u.c. (0.39 nm) corresponding to the unit cell of SRO is observed. However, upon closer inspection of Figure 2(b), it is seen that at certain step edges, the height profile is 0.2 nm, resembling that of a SrO terminated region. The growth of SRO is known to have a preference on the underlying termination in Nb:STO and observed to be faster on TiO$_2$ terminated regions than on SrO termination. Such disparate growth as seen in the AFM image in Figure 2(b) has been marked as Region A and Region B corresponding to trenches along the film plane in SrO terminated regions and terraces (regions in between the trenches) corresponding to TiO$_2$ terminated regions respectively. Further, the growth kinetics changes with increasing thickness of SRO and for a 10 u.c. thin film of SRO we observe closing of the trenches (image not shown) as vicinal surface morphology is obtained. This is in consonance with earlier reported observations.[10] Figure 2(c) shows the XRD pattern of the 10 u.c. SrRuO$_3$ film. Only the *(00h)* diffraction peaks of SrRuO$_3$ is obtained, indicating that the films have a perfect *c*-axis orientation. It is observed that the film is highly strained with the out-of-plane lattice parameter being 0.405 nm.

In order to investigate the electrical transport across a Schottky interface of SRO with Nb:STO, device structures were patterned using standard UV photolithography and Ion Beam Etching (IBE). Ohmic contacts were realized by evaporating Ti/Au to the back of the Nb:STO substrate. Figure 3 shows the room temperature *I-V* characteristics of a 8 u.c. SRO film on Nb:STO. The rectification is as expected for a Schottky diode between a metal and an n-type semiconductor. The linear increase in current with bias as seen in the forward characteristics of the diode can be fitted with the thermionic emission model given by:

$$I = AA^*T^2 \exp\left(-\frac{q\varphi}{k_BT}\right)\left[\exp\left(\frac{qV}{nk_BT}\right) - 1\right] \quad (1)$$

The symbols have their usual meanings.[17] From the linear part of the forward characteristics of the diode, the SBH, *φ*, was determined to be 1.15±0.02 eV with an ideality factor n = 1.1.

This indicates thermionic emission to be the dominant transport mechanism across the Schottky interface.

For the BEEM studies, as shown in Figure 1, a large area contact to the rear of the Nb:STO semiconductor was used to collect the hot electrons after transmission through the SRO film and across the Schottky interface. All the measurements were performed at room temperature (RT). The STM surface topography of a 8 u.c. SRO film, recorded at -1 V and 1 nA is shown in Figure 4. It is observed that the surface topography replicates the AFM image and here too trenches and terraces are clearly visible as shown in A and B respectively. BEEM spectra ($I$-$V$) were obtained by varying the applied bias $V_T$, while keeping the tunnel current ($I_T$) constant using feedback. Each spectrum is a representation of nearly 100 spectra obtained from several similar locations and from different devices. By moving the STM tip, the BEEM spectra were also recorded at different regions on each device viz. the trenches (Region A) and terraces (Region B).

Figure 5 is a representative BEEM spectra of a device with 8 u.c. SRO film on Nb:STO. Figure 5 (a) and (b) corresponds to Region A and Region B respectively. In both cases, we observe a sharp onset in $I_B$ beyond a certain tip bias voltage which corresponds to the local Schottky Barrier Height at the metal-semiconductor interface. It is observed that transmission in Region A is an order of magnitude higher than Region B. This can be understood by the fact that $I_B$ exponentially decreases with increasing thickness of the metal layer, [15] which in Region A is thinner than in Region B (due to reasons discussed in Section 2.2).

The local SBH's at the different regions can be extracted from their respective $I_B$ plots using the Bell-Kaiser model, by plotting the square root of the BEEM current ($I_B$) versus tip bias ($V_T$) using:

$$\sqrt{\frac{I_B}{I_T}} \propto (V_T - \varphi) \tag{2}$$

The intersect of the straight line with the voltage axis gives the local SBH, which for Region A is found to 1.27 ± 0.02 eV and 1.10 ± 0.02 eV for Region B as is shown in Figure 5 (a) and (b) respectively. Thus, we find local differences in the energy band alignments at the M/S interface in Regions A and B. We recall that from standard current-voltage studies, an unique value for the SBH corresponding to 1.15±0.02 eV was obtained (Figure 3), which matches closely to that obtained from Region B using BEEM.

Scanning Tunnelling Spectroscopy (STS) studies were also done at these local regions to gain insights into the local electronic structure of the surface of SRO. Tunneling conductance spectra were recorded by varying $V_T$. The feedback loop of the STM was disengaged and the bias voltage was swept through the desired voltage range to record each spectrum. Upto 200 spectra were recorded each for Region A and B as shown in Figure 6 (a) and (b). From the STS spectra, the terraces are found to be more metallic than the trenches and ascribed to the increased thickness of the SRO at these locations, owing to the favourable underlying $TiO_2$ terminations in the substrate. From earlier reports, it is well known that the metallicity and conductance of SRO films increases with increasing thickness. [18]

To understand the origin of the different transport characteristics in Region A and B, we make an analogy with the bond polarization theory of Tung[19] by invoking the interaction between the chemical bonds with the different dipoles at the local M/S interface. In this model, the SBH is written as

$$\varphi = \gamma_B (\varphi_M - \chi) + (1 - \gamma_B)\frac{E_g}{2} \tag{3}$$

where $\gamma_B$ is the strength of the interface dipole and expressed as

$$\gamma_B = 1 - \frac{q^2 N_B d_{MS}}{\epsilon_{it} (E_g + \kappa)} \tag{4}$$

$N_B$ is the number of bonds at the M/S interface, $E_g$ is the band gap of the semiconductor, $\varepsilon_{it}$ is the dielectric screening at the interface, $d_{MS}$ is the distance between the metal and semiconductor atoms at the interface and $\kappa$ is the sum of all the hopping interactions between neighboring atoms at the interface. Such a model, that takes into account the dipole density at the M/S interface, has been used by Hikita *et al.* [20], along with polarity mismatch at the Schottky interface to explain the systematic increase in the SBH with increasing SrO coverage.

In our experiments, the trenches represent regions with underlying SrO termination as discussed earlier. Here the larger BEEM transmission is also accompanied by a higher local SBH. At the terraces (i.e regions in between the trenches) the BEEM transmission is lower due to the increased thickness of SRO, which grows favorably on underlying TiO$_2$ terminated regions, with a concomitant decrease in the SBH. These differences in the SBH indicate local differences in the band alignments at the M/S interface. We know that in SRO, the Ru *4d* states are split by an octahedral crystal field into $t_{2g}$ and $e_g$ bands and the latter are relevant for hot electron transport in SRO.[21] The *4d$_{3z2-r2}$* orbitals in the $e_g$ bands have a different spatial distribution as compared to the *d$_{x2-y2}$* orbitals;[22] with the orientation of the former being along the interface and the latter parallel to the interface. This difference in spatial distribution and the local strain will influence their interaction with the interface dipoles at the local SrO and TiO$_2$ terminated regions in Nb:STO, with the dipole strength decreasing at the SrO terminations and giving rise to a higher SBH (as also expected from eq.3 and 4). At the TiO$_2$ terminated regions, the strong interaction with the *4d$_{3z2-r2}$* orbitals oriented along the interface will increase the interface dipole strength and decrease the SBH. The strong influence of the underlying substrate termination to the growth of SRO thus gets manifested as local changes in the transport characteristics and their corresponding energy band lineups at the M/S interface which has been probed using the local probing capabilities in BEEM.

The high selectivity of SRO growth with regard to the different surface terminations at local regions of the substrate leads to local variations in the thickness of SRO which is studied using STM and STS. Using the local probing capabilities of BEEM we can capture the differences in the transport and the energy band lineups at such local regions with high spatial resolution. We ascribe this to the differences in the nature and spatial distribution of the interface dipoles at the local M/S interfaces. The ability to detect such changes in transport in buried M/S interfaces using BEEM is a new approach to probe transport in oxide thin films. This approach can also be extended to investigate the role of strong correlation in other complex oxide heterostructures besides characterising the transport parameters relevant for designing oxide electronic devices.

**Acknowledgements**

We thank B. Noheda and T. T. M. Palstra for use of the Pulsed Laser Deposition system. Technical support from J. Baas and J. G. Holstein is thankfully acknowledged. We also acknowledge useful discussions with Y. Hikita and H. Y. Hwang. This work is supported by the Netherlands Organization for Scientific Research NWO-FOM (*nano*) and VIDI program and the Rosalind Franklin Fellowship.

**References:**

[1] A. Ohtomo, H. Y. Hwang, *Nature (London)* **2004**, *427*, 423.

[2] H. Y. Hwang, Y. Iwasa, M. Kawasaki, B. Keimer, N. Nagaosa, Y. Tokura, *Nature Mater.* **2012**, *11*, 103.

[3] A. P. Ramirez, *Science* **2007**, *315*, 1377.

[4] E. Dagotto, *Science* **2007**, *318*, 1076.

[5] V. Garcia, M. Bibes, L. Bocher, S. Valencia, F. Kronast, A. Crassous, X. Moya, S. Enouz-Vedrenne, A. Gloter, D. Imhoff, C. Deranlot, N. D. Mathur, S. Fusil, K. Bouzehouane, A. Barthélémy, *Science* **2010**, *327*, 1106.

[6] M. Bowen, J. -L. Maurice, A. Barthélémy, M. Bibes, D. Imhoff, V. Bellini, R. Bertacco, D. Wortmann, P. Seneor, E. Jacquet, A. Vaurés, J. Humbert, J.-P. Contour, C. Colliex, S. Blugel, P. H. Dederichs, *J. Phys.:Condens. Matter.* **2007**, *19*, 315208.

[7] J. Siedel, L. W. Martin, Q. He, Q. Zhan, Y. –H. Chu, A. Rother, M. E. Hawkridge, P. Maksymovych, P. Yu, M. Gajek, N. Balke, S. V. Kalinin, S. Gemming, F. Wang, G. Catalan, J. F. Scott, N. A. Spaldin, J. Orenstein, R. Ramesh, *Nature Mater.* **2009**, *8*, 229.

[8] S. Farokhipoor, B. Noheda, *Phys. Rev. Lett*. **2011**, *107*, 127601.

[9] J. M. Longo, P. M. Raccah, J. B. Goodenough, *J. Appl. Phys*. **1968**, *39*, 1327.

[10] G. Koster, L. Klein, W. Siemons, G. Rijnders, J. S. Dodge, C-B, Eom, D. H. A. Blank, M. R. Beasley, *Rev. Mod. Phys*. **2012**, *84*, 253.

[11] R. Bachelet, F. Sanchez, J. Santiso, C. Munuera, C. Ocal, J. Fontcuberta, *Chem. Mater*. **2009**, *21*, 2494.

[12] W. J. Kaiser, L. D. Bell, *Phys. Rev. Lett.* **1988**, 60, 1406.

[13] E. Haq, T. Banerjee, M. H. Siekman, J. C. Lodder, R. Jansen, Appl. Phys. Lett. **2005**, *87*, 182912.

[14] S. Parui, K. G. Rana, L. Bignardi, P. Rudolf, B. J. van Wees, T. Banerjee, *Phys. Rev. B* **2012**, *85*, 235416.

[15] M. K. Weilmeier, W. H. Rippard, R. A. Buhrman*, Phys. Rev. B* **1999**, *59*, R2521.

[16] G. Koster, B. L. Kropman, G. J. H. M. Rijnders, D. H. A. Blank, H. Rogalla, *Appl. Phys. Lett.* **1998,** *73*, 2920.

[17] a) S. M. Sze, *Physics of Semiconductor Devices*, 3rd ed., John Wiley & Sons, NY **1969**. b) E.H. Rhoderick, R.H. Williams, *Metal-Semiconductor Contacts*, 2nd Ed., Clarendon-Oxford, **1988**.

[18] H. Kumigashira, H. M. Minohara, M. Takizawa, A. Fujimori, D.Toyota, I. Ohkubo, M. Oshima, M. Lippmaa, M. Kawasaki, *Appl. Phys. Lett*. **2008**, *92*, 122105.

[19] R. T. Tung, *Phys. Rev. B* **2001**, *64*, 205310.

[20] Y. Hikita, M. Nishikawa, T. Yajima, H. Y Hwang, *Phys. Rev. B* **2009**, *79*, 073101.

[21] a) E. B. Guedes  M. Abbate, K. Ishigami, A. Fujimori, K. Yoshimatsu, H. Kumigashira, M. Oshima, F. C. Vicentin, P. T. Fonseca, R. J. O. Mossanek, *Phys. Rev. B* **2012**, *86*, 235127. b) A. J. Grutter, F. J. Wong,  E. Arenholz, A. Vailionis, Y. Suzuki, *Phys. Rev. B* **2012**, *85*, 134429.

[22] D. Pesquera, G. Herranz, A. Barla, E. Pellegrin, F. Bondino, E. Magnano, F. Sánchez, J. Fontcuberta. *Nat. Commun.* **2012**, *3*, 1189.

**Figure Captions:**

**Figure 1.** Schematic diagram of BEEM and energy-band diagram of the BEEM experiment. (a) A Pt-Ir STM tip injects hot electrons over the vacuum barrier into a thin metallic film of $SrRuO_3$. (b) The transmitted electrons that satisfy the energy and momentum criteria to overcome the SBH at the interface are collected in the $Nb:SrTiO_3$ semiconductor (substrate).

**Figure 2.** (a) AFM images (scan area $2\times2\mu m^2$) of chemically terminated substrate with its corresponding step height profile shown in the right along the white line in the image. (b) AFM scan image of a 8 u.c. $SrRuO_3$ film on the substrate. Region A, shown in the image, corresponds to SrO terminated areas and Region B corresponds to $TiO_2$ terminations. $SrRuO_3$ grows more slowly in Region A than in Region B. (c) XRD peaks depicting 2theta-omega scan of 10 u.c. thick $SrRuO_3$ grown on $Nb:SrTiO_3$.

**Figure 3.** Current (I)-Voltage (V) characteristics of the $SrRuO_3/Nb:SrTiO_3$ Schottky diode. The straight lines indicate the thermionic emission theory fitted to obtain the Schottky barrier height and ideality factor of the diode.

**Figure 4.** A STM topography scan of an 8 u.c. thick $SrRuO_3$ film on $Nb:SrTiO_3$ at $V_T = -1$ V and $I_T = 1$ nA. Thermal annealing of the substrate leads to out diffusion of Sr to the surface and finally to step edges. The slower growth of the initial $SrRuO_3$ layer at the SrO terminated areas results in the formation of trenches as shown in Region A. Region B is the area in between the trenches.

**Figure 5.** (a) BEEM transmission per nanoampere of injected tunnel current plotted against $V_T$, at Region A and B for a 8 u.c. $SrRuO_3$ on $Nb:SrTiO_3$. Measurements were done at RT (b) The corresponding SBH extracted from the straight line fit obtained from the Bell Kaiser equation for both Regions A and B.

**Figure 6.** Scanning Tunneling Spectroscopy spectra acquired at Regions A and B at RT by sweeping the applied bias as shown. Region B is found to be more metallic than Region A.

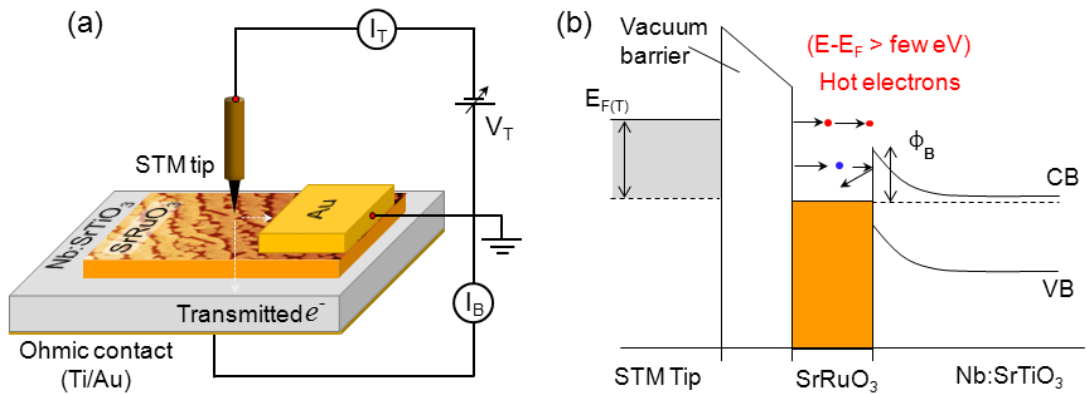

**Figure 1.** Schematic diagram of BEEM and energy-band diagram of the BEEM experiment. (a) A Pt-Ir STM tip injects hot electrons over the vacuum barrier into a thin metallic film of $SrRuO_3$. (b) The transmitted electrons that satisfy the energy and momentum criteria to overcome the SBH at the interface are collected in the $Nb:SrTiO_3$ semiconductor (substrate).

Roy *et al.* Fig. 1

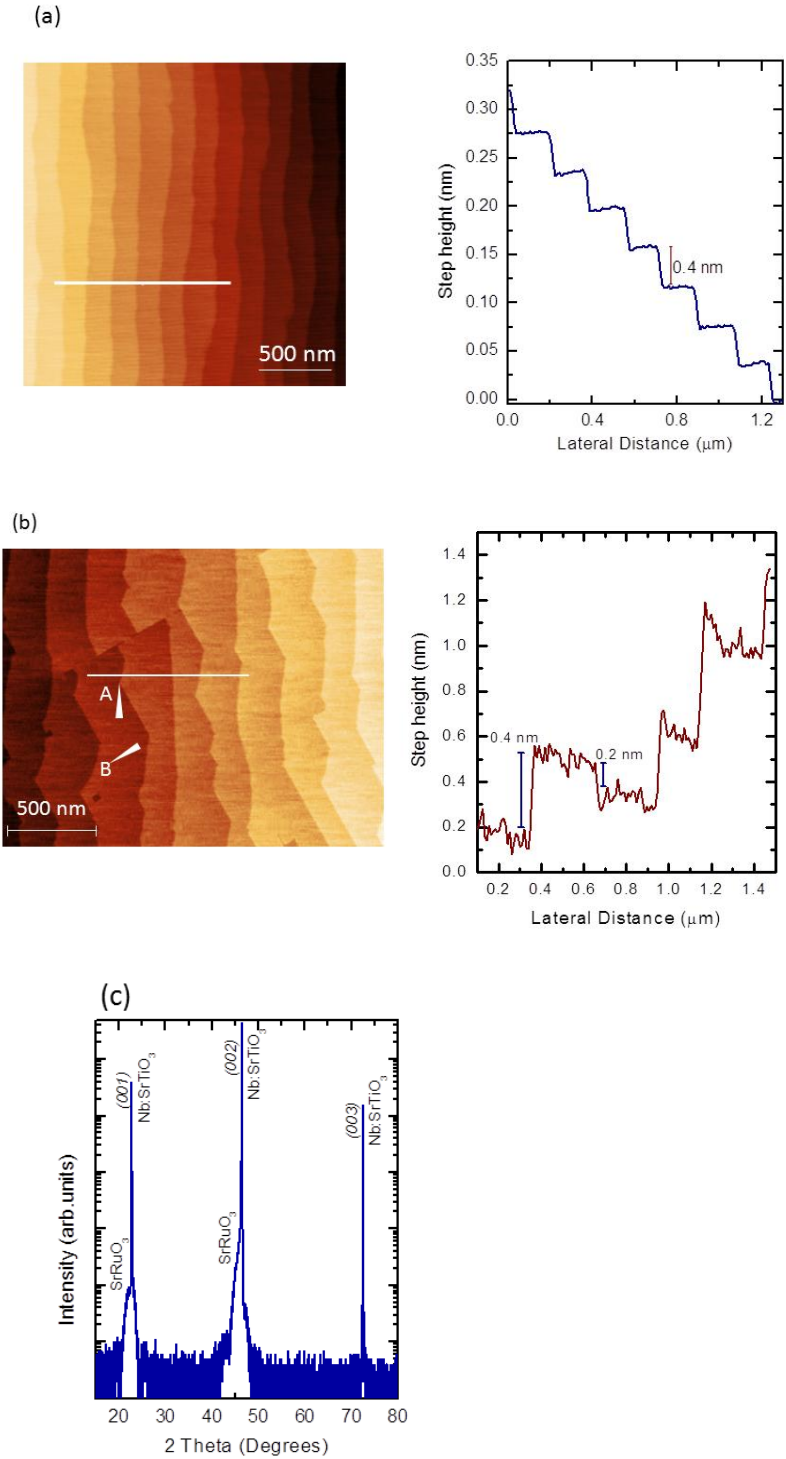

**Figure 2.** (a) AFM images (scan area 2×2μm$^2$) of chemically terminated substrate with its corresponding step height profile shown in the right along the white line in the image. (b) AFM scan image of a 8 u.c. SrRuO$_3$ film on the substrate. Region A, shown in the image, corresponds to SrO terminated areas and Region B corresponds to TiO$_2$ terminations. SrRuO$_3$ grows more slowly in Region A than in Region B. (c) XRD peaks depicting 2theta-omega scan of 10 u.c. thick SrRuO$_3$ grown on Nb:SrTiO$_3$.

Roy *et al.* Fig. 2

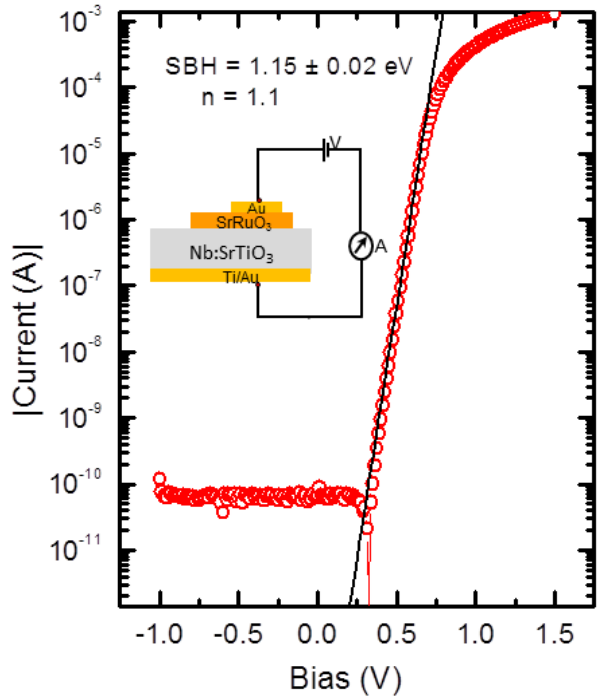

**Figure 3.** Current (I)-Voltage (V) characteristics of the SrRuO$_3$/Nb:SrTiO$_3$ Schottky diode. The straight lines indicate the thermionic emission theory fitted to obtain the Schottky barrier height and ideality factor of the diode.

Roy *et al.* Fig. 3

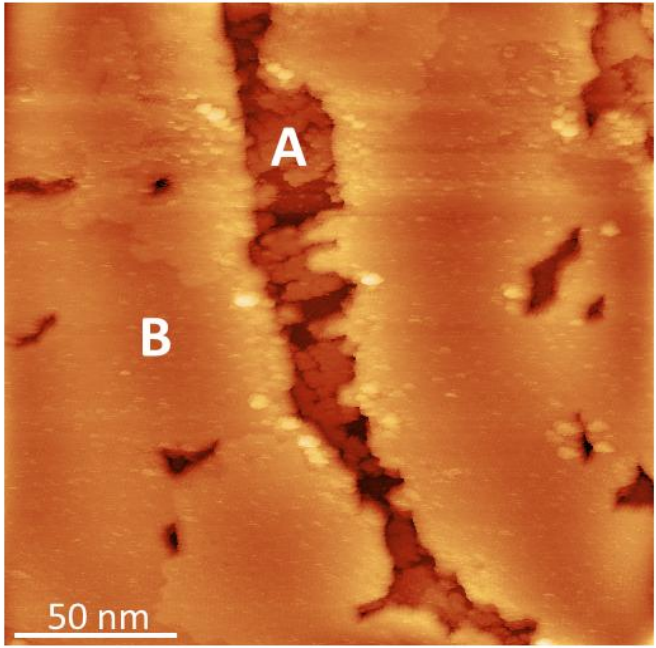

**Figure 4.** A STM topography scan of an 8 u.c. thick SrRuO$_3$ film on Nb:SrTiO$_3$ at $V_T = -1$ V and $I_T = 1$ nA. Thermal annealing of the substrate leads to out diffusion of Sr to the surface and finally to step edges. The slower growth of the initial SrRuO$_3$ layer at the SrO terminated areas results in the formation of trenches as shown in Region A. Region B is the area in between the trenches.

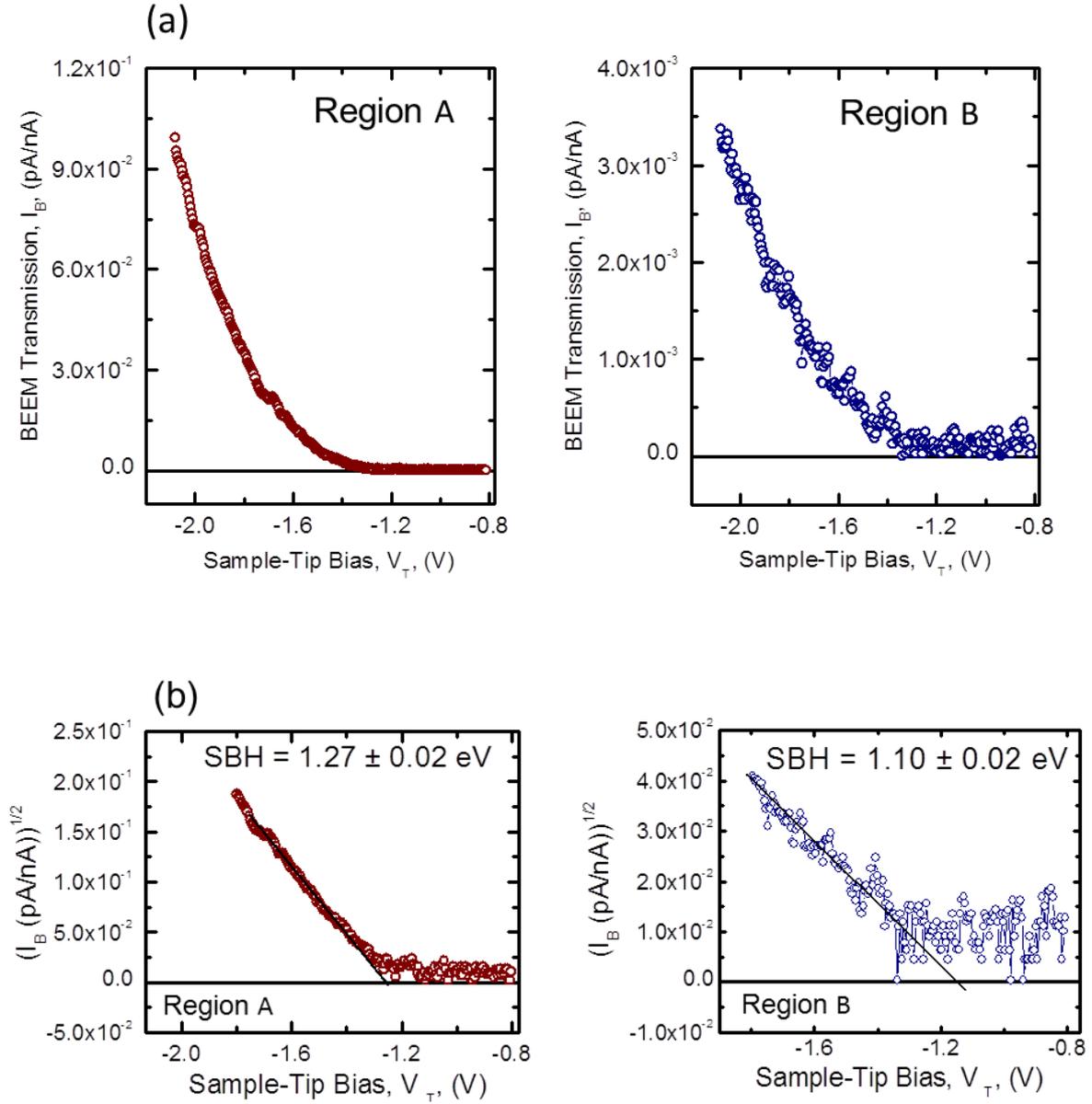

**Figure 5**. (a) BEEM transmission per nanoampere of injected tunnel current plotted against $V_T$, at Region A and B for a 8 u.c. $SrRuO_3$ on $Nb:SrTiO_3$. Measurements were done at RT (b) The corresponding SBH extracted from the straight line fit obtained from the Bell Kaiser equation for both Regions A and B.

Roy *et al.* Fig. 5

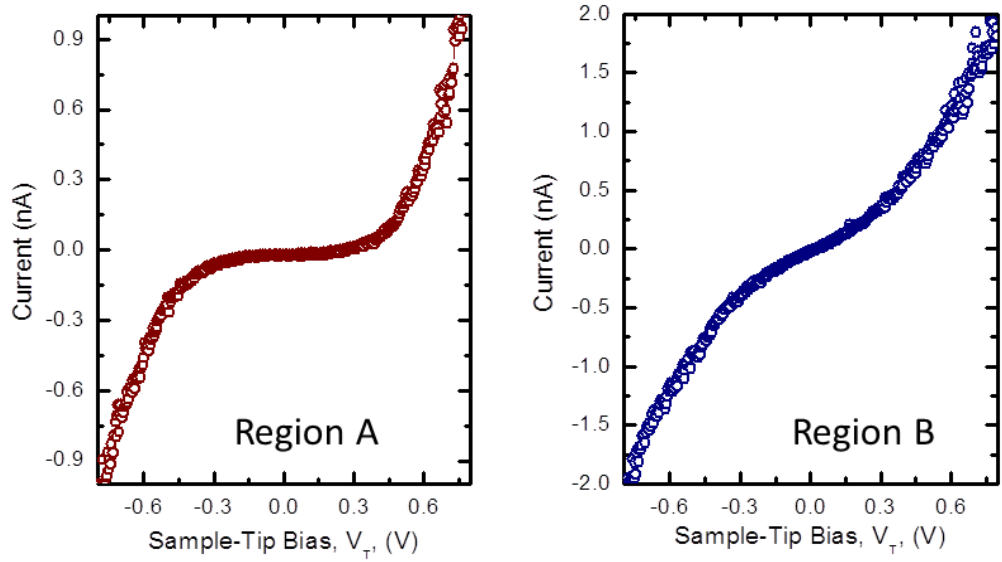

**Figure 6.** Scanning Tunneling Spectroscopy spectra acquired at Regions A and B at RT by sweeping the applied bias as shown. Region B is found to be more metallic than Region A.

Roy *et al.* Fig. 6